\documentclass[10pt,twocolumn]{article}

\usepackage{amsmath}
\usepackage{amssymb}
\usepackage{amsthm}
\usepackage{color}
\usepackage[usenames,dvipsnames,table]{xcolor}
\usepackage[top=54pt, left=54pt, right=54pt, bottom=54pt, paper=letterpaper]{geometry}
\usepackage[table]{xcolor}
\usepackage[noadjust]{cite}
\usepackage{booktabs}
\usepackage[format=plain,labelfont=it]{caption}
\usepackage[colorlinks=false]{hyperref}
\usepackage{url}
\usepackage{algpseudocode}
\usepackage{algorithm}
\usepackage{tabularx}
\usepackage{titlesec}
\usepackage{accents}
\usepackage{graphics}
\usepackage{graphicx}
\usepackage{array}
\usepackage{matlab-prettifier}

\usepackage{enumitem}
\usepackage{relsize}
\usepackage{subcaption}
\usepackage[T1]{fontenc}

\titleformat{\section}{\centering\normalfont\scshape}{\Roman{section}.}{5pt}{}
\titleformat{\subsection}{\normalfont\it}{\Alph{subsection}.}{5pt}{}
\titleformat{\subsubsection}{\normalfont\it}{\hspace{4mm}\arabic{subsubsection})}{5pt}{}

\newcommand\infoFootnote[1]{%
  \begingroup
  \renewcommand\thefootnote{}\footnote{#1}%
  \addtocounter{footnote}{-1}%
  \endgroup}

\usepackage{tikz}
\usetikzlibrary{arrows}
\usetikzlibrary{shapes}
\usetikzlibrary{decorations.pathmorphing}
\usetikzlibrary{decorations.pathreplacing}
\usetikzlibrary{decorations.shapes}
\usetikzlibrary{decorations.text}
\usetikzlibrary{positioning, shapes}

\tikzset{
block/.style = {draw, fill=white, rectangle, minimum height=3em, minimum width=4.5em},
tmp/.style  = {coordinate}, 
sum/.style= {draw, fill=white, circle, node distance=2cm},
input/.style = {coordinate},
output/.style= {coordinate}
pinstyle/.style = {pin edge={to-,thick,black}}
}

\definecolor{matgreen}{RGB}{0, 128, 19}
\lstdefinestyle{matstyle}{
    style = Matlab-editor,
    basicstyle = \small\ttfamily,
    escapechar=`
}
\lstdefinestyle{smallerstyle}{
    style = Matlab-editor,
    basicstyle = \fontsize{8.5}{10}\ttfamily,
    escapechar=`
}

\newcommand{\mattext}[1]{\texttt{\fontsize{9.5}{12}\selectfont #1}}

\newcommand{\R}{\mathbb{R}} 
\newcommand{\N}{\mathbb{N}} 

\newcommand{\Z}{\mathbb{Z}}

\newcommand{\Rc}{\mathcal{R}}
\newcommand{\Zc}{\mathcal{Z}}

\newcommand{\ab}{\boldsymbol{a}}

\newcommand{\yb}{\boldsymbol{y}}

\newcommand{\xb}{\boldsymbol{x}}

\newcommand{\ub}{\boldsymbol{u}}
\newcommand{\zb}{\boldsymbol{z}}

\newcommand{\Ab}{\boldsymbol{A}}
\newcommand{\Bb}{\boldsymbol{B}}
\newcommand{\Cb}{\boldsymbol{C}}

\newcommand{\Ib}{\boldsymbol{I}}

\newcommand{\Db}{\boldsymbol{D}}

\newcommand{\Kb}{\boldsymbol{K}}

\newcommand{\ct}{\mathsf{ct}}
\newcommand{\sk}{\mathsf{sk}}
\newcommand{\pk}{\mathsf{pk}}
\newcommand{\ek}{\mathsf{ek}}
\newcommand{\evk}{\mathsf{evk}}

\newcommand{\skb}{\boldsymbol{\mathrm{sk}}}

\newcommand{\zerob}{\boldsymbol{0}}

\newcommand{\modq}{\;\mathrm{mod}\,q}

\renewcommand{\mod}[1]{\,\mathrm{mod}\,#1}

\newcommand{\Enc}{\mathsf{Enc}}
\newcommand{\Dec}{\mathsf{Dec}}
\newcommand{\Ecd}{\mathsf{Ecd}}
\newcommand{\Dcd}{\mathsf{Dcd}}

\newcommand{\LWE}{\mathsf{LWE}}
\newcommand{\GSW}{\mathsf{GSW}}

\newcommand{\round}[1]{\left\lfloor#1\right\rceil}

\newcommand{\floor}[1]{\left\lfloor#1\right\rfloor}

\newcommand{\blind}[1]{{\color{white}{#1}}}

\title{\vspace{-2mm}\bf
A code-driven tutorial on encrypted control:\\ From pioneering realizations to modern implementations$^\ast$}
\author{Nils Schl\"uter, Junsoo Kim, and Moritz Schulze Darup\vspace{2mm}}
\date{}

\author{Nils Schl\"uter, Junsoo Kim, and Moritz Schulze Darup}

\begin{document}
\maketitle

% Slightly reduce vertical padding around displaystyle math
\setlength{\abovedisplayskip}{8pt plus 2pt minus 1pt}  % default: 10.0pt plus 2.0pt minus 5.0pt
\setlength{\belowdisplayskip}{8pt plus 2pt minus 1pt}  % default: 10.0pt plus 2.0pt minus 5.0pt

\textbf{\textit{Abstract}.} {\bf%
The growing interconnectivity in control systems due to robust wireless communication and cloud usage paves the way for exciting new opportunities such as data-driven control and service-based decision-making.
At the same time, connected systems are susceptible to cyberattacks and data leakages. 
Against this background, encrypted control aims to increase the security and safety of cyber-physical systems.
A central goal is to ensure confidentiality of process data during networked controller evaluations, which is enabled by, e.g., homomorphic encryption.
However, the integration of advanced cryptographic systems renders the design of encrypted controllers an interdisciplinary challenge. 

This code-driven tutorial paper aims to facilitate the access to encrypted control by providing exemplary realizations based on popular homomorphic cryptosystems.
In particular, we discuss the encrypted implementation of state feedback and PI controllers using the Paillier, GSW, and CKKS cryptosystem.
\infoFootnote{N. Schl\"uter and M.~Schulze Darup are with the Control and Cyberphysical Systems Group, Dept.~of Mechanical Engineering, TU Dortmund University, Germany.
        {E-mails:  \{nils.schlueter, moritz.schulzedarup\}@tu-dortmund.de}.}
\infoFootnote{J. Kim is with the Department of Electrical and Information Engineering, Seoul National University of Science and Technology, South Korea.
E-mail: junsookim@seoultech.ac.kr}
\infoFootnote{Financial support by the German Research Foundation (DFG) under the grant SCHU 2940/4-1 is gratefully acknowledged. \vspace{0.5mm}}
\infoFootnote{\hspace{-1.5mm}$^\ast$This paper is a \textbf{preprint} of a contribution to the 22nd European Control Conference 2024.}%
}

\noindent

\section{Introduction}
Safety in terms of stability, robustness, and resilience is one of the most influential concepts in classical control. Clearly, for cyber-physical systems (CPS), safety shall also take the vulnerability to cyberattacks into account. 
In this context, a characteristic property of networked control systems is that data is processed ``externally'' 
on possibly untrustworthy platforms such as cloud servers or neighboring agents. Consequently, the security of sensitive process data is threatened, and a holistic approach is needed to ensure, e.g., confidentiality throughout the control loop. 

Encrypted control (see \cite{darup2021encrypted,Schlueter2023} for surveys) addresses this challenge by implementing and evaluating networked controllers in an encrypted fashion.
In this context, a pivotal technology is homomorphic encryption
(HE, see \cite{marcolla2022survey} for an introduction), 
which enables computations on encrypted data. 
Yet, achieving an efficient integration of control and HE poses non-trivial interdisciplinary challenges. Intensifying research efforts is essential for addressing these challenges.
Thus, the tutorial paper aims to facilitate access to encrypted control for researchers lacking a cryptographic background. To this end, we provide crucial insights on the commonly used Paillier \cite{Paillier1999}, Gentry-Sahai-Waters (GSW) \cite{gentry2013homomorphic}, and Cheon-Kim-Kim-Song (CKKS) \cite{Cheon2017_CKKS} cryptosystems, and illustrate how these schemes enable the realization of basic encrypted controllers, spanning from pioneering to modern implementations. 
To enhance the tutorial aspect, we provide code for each controller realization.
For instance, we will explain how linear state feedback of the form
\begin{equation}
\label{eq:uKx}
    \ub(k)=\Kb \xb(k)
\end{equation}
can be evaluated based on encrypted states $\xb$ and encrypted controller parameters $\Kb$ (without intermediate decryptions).

The remaining paper is organized as follows. In Section~\ref{sec:basics}, we provide a brief overview of HE in general and the three cryptosystems Paillier, GSW, and CKKS used for the upcoming implementations. Section \ref{sec:controllers} discusses necessary controller reformulations for an encrypted realization of linear state feedback and PI control.
Actual implementations of the two control schemes using a custom Matlab toolbox developed by the authors are presented in Section~\ref{sec:implementation} followed by numerical experiments in Section~\ref{sec:numericalStudies}.
Finally, a summary of the tutorial paper and an outlook are given in Section \ref{sec:outlook}.

\textit{Notation.} The sets of real, integer, and natural numbers are denoted by $\R$, $\Z$, and $\N$, respectively. 
Central to this work is the set $\Z/q\Z$ of residue classes of the integers modulo $q\in\N$ with $q>1$.
In this regard, for $z\in\Z$, the modulo reduction $z \mod q := z - m q$ refers to a mapping from $\Z$ onto a representative of $\Z/q\Z$ specified by $m$. 
We frequently use the representatives $\Z_q:=\{0,1,\ldots,q-1\}$ (where $m := \floor{z/q}$) and $\Zc_q:=[-q/2,q/2)\cap \Z$ (where $m:=\round{z/q}$). 
Here, $\floor{\cdot}$ and $\round{\cdot}$ are the (element-wise) flooring and rounding functions, respectively.
Finally, $a = b \modq$ indicates a congruence relation.

\section{An overview of homomorphic cryptosystems}
\label{sec:basics}

Cryptosystems come with at least three algorithms, i.e., an encryption $\Enc_{\ek}()$, a decryption $\Dec_{\sk}()$, and a key generation $\mathsf{KeyGen()}$, which outputs the key(s) required for encryption and decryption.
The encryption takes a message~$z$ from the cryptosystem's message space and the ``encryption key'' $\ek$ and outputs a ciphertext $\ct(z)=\Enc_{\ek}(z)$ that lies in the cryptosystem's ciphertext space. The security of the cryptosystem then guarantees that an attacker cannot infer information about $z$ given $\ct(z)$. A decryption can, however, be carried out via $\Dec_{\sk}(\ct(z))$ using the secret key $\sk$. 
Now, $\ek$ either equals the secret key $\sk$ or refers to a public key $\pk$. If $\ek=\sk$, then the scheme is called symmetric because $\sk$ is used for both encryption and decryption. 
In contrast, in an asymmetric (or public-key) scheme,
encryptions can be carried out with $\pk$, and only the decryption requires $\sk$. Hence, $\pk$ can be distributed without the risk of eavesdropping (see \cite{katz2020introduction} for further details).

Now, HE extends these algorithms by encrypted computations that are enabled by 
algebraic structures preserved under the encryption via $\Enc_{\ek}()$. 
More formally, additively HE schemes offer an operation ``$\oplus$'' which allows the evaluation of encrypted additions according to
\begin{equation}
    \label{eq:addhom}
    \ct(z_1 + z_2) = \ct(z_1) \oplus \ct(z_2).
\end{equation}
Analogously, multiplicatively HE schemes support encrypted multiplications ``$\otimes$'' via
\begin{equation}
    \label{eq:multhom}
    \ct(z_1 z_2) = \ct(z_1) \otimes \ct(z_2).
\end{equation}
Depending on what operations are provided, HE schemes are categorized as follows.
\begin{itemize}%[leftmargin=0pt]
    \item \textit{Partially homomorphic encryption} (PHE) allows either for \eqref{eq:addhom} or \eqref{eq:multhom}, but not both.
    \item \textit{Fully homomorphic encryption} (FHE) allows for an unlimited amount of operations \eqref{eq:addhom} and \eqref{eq:multhom}. 
    \item \textit{Leveled FHE} allows for a finite amount of \eqref{eq:addhom} and \eqref{eq:multhom}.
\end{itemize}
Remarkably, the denomination ``fully'' stems from the fact that 
\eqref{eq:addhom} and \eqref{eq:multhom} 
suffice to compute (in principle) any function.
Furthermore, a ``level''  essentially refers to the maximum amount of multiplications one can perform on a ciphertext with a leveled FHE scheme. Conceptually, each multiplication consumes a level until zero is reached. 
If a level is not sufficient to perform certain computations, enabling additional encrypted multiplications requires a computationally highly demanding routine called bootstrapping (which essentially resets the level) \cite{cheon2018bootstrapping}. For real-time critical control systems, bootstrapping is typically not an option. Hence, knowing the ``multiplicative depth'' (cf.~Fig.~\ref{fig:CircuitKy}) of the computations to be carried out and choosing a suitable cryptosystem is often crucial.
To get a better feeling for such a choice, we briefly summarize the popular HE schemes Paillier, GSW, and, CKKS next.
We stress, at this point, that a detailed understanding of the cryptosystems is not required to realize encrypted controllers. However, aiming for efficient implementations, it is helpful to be aware of some fundamental characteristics.

\subsection{PHE via the Paillier scheme} 
\label{subsec:paillier}

Paillier \cite{Paillier1999} 
is a public-key scheme, which offers PHE in terms of \eqref{eq:addhom}.
To set it up,  two large primes $p_1,p_2\in\N$ with bit lengths of $\lambda$ are selected during $\mathsf{KeyGen}(\lambda)$. The public and secret key are then specified as $\pk = p_1 p_2$ and ${\sk = {(p_1-1)}(p_2-1)}$, respectively. 
Next, for a message $z\in\Z_{\pk}$, encryption and decryption is carried out via
\begin{align}
\label{eq:paillierencdec}
    \ct(z) &= \Enc_{\pk}(z) = (\pk+1)^z r^{\pk} \mod{\pk^2} \\
    \nonumber
    z &= \Dec_{\sk}(\ct(z)) = \frac{\left(\ct^\sk \mod{\pk^2}\right) - 1}{\pk}\, \sk^{-1} \mod{\pk},
\end{align}
where $r$ is sampled uniformly at random from $\Z_{\pk}$ but 
such that $\mathrm{gcd}(r,\pk)=1$.
One can then readily show that the homomorphism~\eqref{eq:addhom} is indeed given by
\begin{equation}
\label{eq:paillierhom}
\ct(z_1)\oplus \ct(z_2) = \ct(z_1) \ct(z_2) \mod{\pk^2}.
\end{equation}
Notably, 
Paillier offers another homomorphism in the form of
\begin{equation}
\label{eq:paillierMul}
\ct(z_1 z_2) = \ct(z_1)\odot z_2 =  \ct(z_1)^{z_2} \mod{\pk^2},
\end{equation}
i.e., a partially encrypted multiplication with public factors  $z_2\in\Z_{\pk}$.
We finally note that the security of the Paillier scheme builds on the 
decisional composite residuosity assumption (and factoring of $\pk$) for a suitable choice of its security parameter $\lambda$, which also determines the size of the message space.

\subsection{Leveled FHE via GSW}
\label{subsec:GSW}

In contrast to Paillier, GSW \cite{gentry2013homomorphic} builds on the so-called learning with errors (LWE) problem \cite{regev2005lattices}, which is well-suited for constructing homomorphic cryptosystems due to its structural simplicity. 
In order to use the GSW scheme, one first performs an additional step by calling $\mathsf{Setup}(\lambda,q)$ which specifies, e.g, the required key dimension $N$ and error distributions in such a way that a security of $\lambda$ bits can be ensured for the message space $\Zc_q$ (of size $q$).
Second, within $\mathsf{KeyGen}()$, one samples a secret key $\skb$ uniformly at random from $\Zc_q^N$. Note that, for simplicity, we here  
focus on a symmetric variant, i.e., no public key is generated. 

Now, the GSW scheme is based on two ciphertext types. Given a message $z\in\Zc_q$, we denote them by $\LWE(z)$ and $\GSW(z)$, respectively. 
In the former case, we can encrypt and decrypt\footnote{Evidently, the decryption returns the message perturbed by the small error $e$. If this is undesired, $\LWE(\delta z)$ with $\delta \in\N$ can be used. The decryption is then extended by $\round{(b+\ab^\top \skb \mod{q})/\delta}$, which results in $z$ as long as $\delta z\in\Zc_q$ and $\delta >2|e|$.} via
\begin{subequations}
    \label{eq:LWEencdec}
    \begin{align}
    \LWE(z) &= \Enc_{\sk}(z) = (b, \ab^\top) + (z,\zerob^\top) \mod{q} \\
    z+e &= \Dec_{\sk}(\LWE(z)) = b+\ab^\top \skb \mod{q},
\end{align}
\end{subequations}
where $b = -\ab^\top \skb + e \mod{q}$, $\ab$ must be sampled uniformly at random from $\Zc_q^{N}$, and the small error $e$ is typically sampled from a discrete Gaussian distribution.

Due to their linearity, an encrypted addition $\LWE(z_1) \oplus \LWE(z_2)$ is realized by (literally) adding LWE ciphertexts, i.e, $\LWE(z_1+z_2)$ is obtained by
\begin{equation}
    \label{eq:LWEadd}
    \LWE(z_1) \oplus \LWE(z_2) = ( b_1+b_2, \ab^\top_1+\ab^\top_2 ) \modq.
\end{equation}
A characteristic property of LWE-based schemes is that the ciphertext error grows with each homomorphic operation. For instance, the error of $\LWE(z_1+z_2)$ is $e_1+e_2$. 
To realize encrypted multiplications, GSW ciphertexts are required, which make clever use of a base-decomposition technique. Essentially, this tames the error term that would otherwise drastically increase and destroy the message. GSW ciphertexts then enable
\begin{subequations}    
\begin{align}
    \label{eq:LWEtoGSW}
    \otimes_{\LWE}:\quad& \GSW \times \LWE \, \to \LWE \\ 
    \label{eq:GSWtoGSW}
    \otimes_{\GSW}:\quad & \GSW \times \GSW \to \GSW
\end{align}
\end{subequations}
but also support encrypted additions by $\GSW(z_1) + \GSW(z_2)$, as in \eqref{eq:LWEadd}. 
Due to space limitations, we omit to specify GSW ciphertexts and the concrete realizations of the two multiplications.
Details can, however, be found in \cite{gentry2013homomorphic} or \cite{kim2020comprehensive}, where the latter is tailored for control engineers.

One advantage of the GSW scheme is that during encrypted computations, the error grows only by a small polynomial factor in contrast to, e.g., CKKS described next. A drawback of the scheme is its computational complexity
proportional to
the square of the key length.

\subsection{Leveled FHE via CKKS}
\label{subsec:CKKS}

CKKS \cite{Cheon2017_CKKS} builds on a ring-variant of the LWE. More precisely, the considered ring-LWE (RLWE) problem results from defining LWE over a quotient ring $\Rc_q$.
Consequently, CKKS operates over polynomials of the form $z(X)=\tilde{z}_{N-1} X^{N-1}+...+\tilde{z}_1 X^1+\tilde{z}_0X^0$, where the degree is strictly less than $N$ and where each coefficient $\tilde{z}_i$ is an element of $\Zc_q$. This entails efficiency benefits in terms of computational complexity and ciphertext size for a given message. 
Nonetheless, CKKS has several commonalities with GSW.
For instance, $\mathsf{Setup}(\lambda,q)$ and $\mathsf{KeyGen}()$ work similarly. 
Furthermore, for a given message $z\in\Rc_q$, CKKS performs (in its symmetric variant) encryption and decryption by means of 
\begin{align*}
    \ct(z)=(b,a)+(z,0)\in\Rc_q^2 \,\,\,\, \text{and} \,\,\,\,  b+a\,\sk = z+e\in\Rc_q,
\end{align*}
where $b = -a\,\sk + e\in\Rc_q$ similarly to \eqref{eq:LWEencdec}. Analogously, ciphertext addition is supported via~\eqref{eq:LWEadd}.

However, there are also important differences between CKKS and GSW.
For instance, instead of a base-decomposition as in GSW, CKKS builds its multiplication on the observation that $(b_1 + a_1 \sk)(b_2 + a_2 \sk)\in\Rc_q$ is an encryption of $z_1 z_2$ and replaces the resulting $\sk^2$ term in a corresponding ciphertext with the help of a special ``evaluation key'' which is generated during $\mathsf{KeyGen()}$. 
Also, the message polynomial $z\in\Rc_q$ has $N$ coefficient slots where up to $N/2$ coefficients can be used by a special encoding. 
Describing CKKS in more detail is intractable here due to space limitations and since the technicalities are beyond the scope of this paper.
However, we refer to \cite{Schlueter2023} for a detailed introduction for engineers.

\section{Reformulations for encrypted control}
\label{sec:controllers}

As specified in \cite[Sect.~2.1.3]{Schlueter2023}, realizing encrypted controllers via HE typically requires integer-based controller reformulations. In the following, we illustrate the underlying process for linear state feedback and a PI controller. We then apply the homomorphic operations summarized in Section~\ref{sec:basics} to derive an encrypted implementation.

\subsection{Data Encoding}
\label{subsec:encoding}

As we have seen, the message and ciphertext spaces of the Paillier, GSW, and (to some extent) CKKS cryptosystem are finite sets, such as $\Z_q$. 
Thus, in order to use them for real-valued data, an \textit{integer-based} encoding $\Ecd()$ and decoding $\Dcd()$ is required. 
For efficiency reasons, we will build on a (generalized) fixed-point encoding of $x\in\R$ here. To this end, we specify a (public) scaling factor $s\geq 1$ and define
\begin{equation}
\label{eq:encoding}
     z = \Ecd_q(x,s) =\round{s x} \modq ,
\end{equation}
where we obtain $z\in\Z_q$ as desired.
Clearly, due to the rounding, $z$ encodes only an approximation of $x$. In case $\Z_q$ is used, the decoding can be achieved via
\begin{equation}
\label{eq:decoding}
x \approx \Dcd_q(z,s) = \begin{cases} z/s & \text{if } z<q/2, \\
(z-q)/s & \text{otherwise.}
\end{cases}
\end{equation}
Here, $\Dcd_q(z,s)$ implements a partial inverse of the modulo operation and then \textit{rescales} the result. 
In case $\Zc_q$ is used, \eqref{eq:encoding} stays the same while the decoding is simplified to $\Dcd_q(z,s) = z/s$.
The reason for this difference is that the modulo reduction for $\Z_q$ maps negative numbers to  $[\floor{q/2},q-1]\cap \Z$. 
Regardless of the representative being used, we find $x \approx \Dcd_q(z,s)$ for $z$ as in~\eqref{eq:encoding} and all $x$ satisfying $\round{sx}\in \Zc_q$. 
More precisely, for all such $x$, we find the (quantization) error $|x-\Dcd_q(z,s)| \leq 1/(2s)$. Importantly, as a result of an overflow, this error becomes very large (at least $q/s$) whenever $\round{sx} \notin \Zc_q$. 

Unfortunately, the standard encoding strategy in CKKS is significantly more complex. Hence, we will simply use \eqref{eq:encoding} element-wise with a suitable padding (details follow later). The interested reader may find a simple introduction in \cite{Schlueter2023} and technical details in \cite{Cheon2017_CKKS,lyubashevsky2013toolkit}.

\subsection{Benchmark controllers}
\label{subsec:benchmarkcontrollers}

Various control schemes have already been realized in an encrypted fashion. Among these are, for example, linear (state or output) feedback \cite{Kogiso2015,Farokhi2016}, linear dynamic control \cite{kim2021dynamic}, polynomial control \cite{schulze2020encrypted}, and model predictive control \cite{SchulzeDarup2018_LCSS,Alexandru2018_CDC}. Often, tailored reformulations are required in order to support the application of homomorphic cryptosystems, or they contribute to a higher efficiency. Here, we focus on introductory realizations of encrypted controllers and thus consider controller types, where complex reformulations are not necessary.
One of the simplest cases then is linear state feedback as in~\eqref{eq:uKx}
with $\Kb \in \R^{m \times n}$, $\ub(k)$ being the control input, and $\xb(k)$ the system's state. 
In fact, after considering an
integer-based approximation
\begin{equation}
\label{eq:uKyInteger}
\ub(k) \approx \frac{1}{s^2} \lfloor s \Kb \rceil \lfloor s \xb(k) \rceil 
\end{equation}
for some sufficiently large $s$, an encrypted realization using the homomorphisms from Section~\ref{sec:basics} is straightforward. 
In fact, as detailed below, 
an encrypted evaluation of the $i$-th control action (i.e., $\ub_i$) can be carried out via
\begin{align}
\label{eq:partiallyEncryptedKy}
\lfloor s \Kb_{i1}\rceil & \! \odot  \ct(\lfloor s \xb_{1} \! \rceil )  \! \oplus ...  \oplus \lfloor s \Kb_{in} \rceil \! \odot \ct( \lfloor s \xb_{n} \rceil) \,\,\text{or} \\
\label{eq:fullyEncryptedKy}
\!\!\ct( \lfloor s \Kb_{i1}\! \rceil) & \! \otimes  \ct(\lfloor  s \xb_{1}\!\rceil )  \!\oplus ... \oplus \ct( \lfloor s \Kb_{in} \!\rceil) \!\otimes \ct(\lfloor s \xb_{n} \!\rceil ),   \!\!
\end{align}
respectively (where ``multiplication before addition'' applies). Clearly, the realizations differ in that the controller coefficients are accessible in~\eqref{eq:partiallyEncryptedKy} whereas \eqref{eq:fullyEncryptedKy} is fully encrypted. 
Note that the rescaling, i.e., the division by $s^2$ in~\eqref{eq:uKyInteger}, is carried out at the actuator after decrypting the encrypted control actions and before applying them. 

\begin{figure}[h]
    \centering
     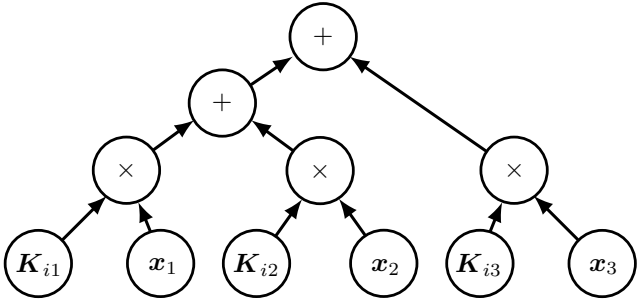
     %}
     \vspace{-4mm} 
    \caption{Arithmetic circuit for the evaluation of $\sum_{j=1}^3 \Kb_{ij} \xb_j$. Note that multiplications only occur once on every directed path, resulting in a multiplicative depth of one.}
    \label{fig:CircuitKy}
    \vspace{-4mm}
\end{figure}

State feedback as in \eqref{eq:uKx} (and, analogously output feedback)
can be easily encrypted mainly for three reasons: There are no iterations involved (for instance, $\ub(k)$ does not depend on $\xb(k-1)$), the multiplicative depth of the corresponding arithmetic circuit is only one (see Fig.~\ref{fig:CircuitKy}), and
$\lfloor s \Kb \rceil \lfloor s \xb(k) \rceil$ 
can be easily bounded (in order to avoid critical overflows during encrypted computations). 
A controller type, where these beneficial features are not immediately given, is linear dynamic control of the form
\begin{subequations}
\label{eq:linearDynamic}
\begin{align}
\xb_c(k+1) &=\Ab_c \xb_c(k) + \Bb_c \yb(k) \\
\ub(k) &= \Cb_c \xb_c(k) + \Db_c \yb(k),
\end{align}
\end{subequations}
where $\yb\in \R^l$ is the system's output, $\xb_c$ is the controller state, and $\Ab_c$, $\Bb_c$, $\Cb_c$, and $\Db_c$ reflect controller parameters.

In particular, the iterative nature of the controller is problematic, as apparent from the integer-based representation
\begin{align*}
\zb(k+1) &= \lfloor s \Ab_c \rceil \zb(k) + \lfloor s \Bb_c \rceil  \lfloor s^{k+1} \yb(k)  \rceil\\
\ub(k) &\approx \frac{1}{s^{k+2}} \left(\lfloor s \Cb_c \rceil \zb(k) + \lfloor s \Db_c \rceil  \lfloor s^{k+1} \yb(k)  \rceil\right)
\end{align*}
with $\zb(0):=\lfloor s \xb_c(0) \rceil$. In fact, in order to ensure that the accumulated scalings in the integer state $\zb(k)$ are matched by the controller inputs $\yb(k)$, we use 
time-varying and increasing scalings in $\lfloor s^{k+1} \yb(k)  \rceil$ (see \cite[Sect. II.B]{SchlueterDarupECC2021} for details). As a consequence, ensuring bounds on $\lfloor s \Cb_c \rceil \zb(k) + \lfloor s \Db_c \rceil  \lfloor s^{k+1} \yb(k)  \rceil$ for all $k\in\N$ is typically unattainable, leading to the risk of overflows. As pointed out in \cite{Cheon2018need}, this issue can be avoided for (almost) integer $\Ab_c$. In fact, a meaningful integer-representation of~\eqref{eq:linearDynamic} then is, e.g., 
\begin{subequations}
\begin{align}
\label{eq:zDynamicsIntegerAc}
\zb(k+1) &= \lfloor \Ab_c \rceil \zb(k) + \lfloor s \Bb_c \rceil  \lfloor s \yb(k)  \rceil\\
\ub(k) &\approx \frac{1}{s^3} \left(\lfloor s \Cb_c \rceil \zb(k) + \lfloor s^2 \Db_c \rceil  \lfloor s \yb(k)  \rceil\right)
\end{align}
\end{subequations}
with $\zb(0):=\lfloor s^2 \xb_c(0) \rceil$. Apart from the design of (almost) integer $\Ab_c$, 
a novel issue is now that such $\Ab_c$
often lead to unstable controller dynamics in~\eqref{eq:zDynamicsIntegerAc}, which is undesirable for networked control~\cite{schlueter2021stability}.
Still, at least marginal stability can be obtained for, e.g., the special case of decoupled proportional-integral (PI) control.
Under the simplifying assumption of constant reference values $\yb_{\text{ref}}(k)=\zerob$) a PI controller is given by $\Ab_c=\Ib_l$, ${\Bb_c=\Delta t \Ib_l}$, $\Cb_c=\Kb_{\text{i}}$, and $\Db_c=\Kb_{\text{p}}$ with the sampling interval $\Delta t>0$ and the diagonal matrices $\Kb_{\text{p}}$ and $\Kb_{\text{i}}$ of dimension $\R^{l \times l}$.
Due to the diagonal controller matrices, the individual controller states $\xb_{c,i}$ and control actions $\ub_i$ are independent by construction. Hence, we can simplify the presentation without loss of generality and consider
a single PI-controlled loop leading to the integer-based representation
\begin{subequations}
\label{eq:scalarPI}
\begin{align}
\label{eq:recursionPI}
z(k+1) &= z(k) + \lfloor s \Delta t \rceil  \lfloor s y(k)  \rceil\\
u(k) &\approx \frac{1}{s^3} \left(\lfloor s K_{\text{i}} \rceil z(k) + \lfloor s^2 K_{\text{p}} \rceil  \lfloor s y(k)  \rceil\right)
\end{align}
\end{subequations}
with scalar controller inputs, states, and outputs.
Analogously to~\eqref{eq:partiallyEncryptedKy} and \eqref{eq:fullyEncryptedKy}, \eqref{eq:scalarPI} can now be implemented partially or fully encrypted.
More precisely, to compute encrypted representatives of the control actions, a cloud will either compute
\begin{align}
\label{eq:partiallyEncryptedDynamicU}
\lfloor s K_{\text{i}} \rceil &\odot  \ct (z(k)) \oplus  \lfloor s^2 K_{\text{p}} \rceil \odot \ct( \lfloor s y(k)  \rceil) \quad \text{or} \\
\label{eq:fullyEncryptedDynamicU}
\ct(\lfloor s K_{\text{i}} \rceil) &\otimes  \ct (z(k)) \oplus  \ct(\lfloor s^2 K_{\text{p}} \rceil) \otimes \ct( \lfloor s y(k)  \rceil). 
\end{align}
Afterwards, encrypted controller state updates are either performed via
\begin{align}
\label{eq:partiallyEncryptedDynamicZ}
\ct(z(k+1)) &= \ct(z(k)) \oplus \lfloor s \Delta t \rceil \odot \ct( \lfloor s y(k)  \rceil) \quad \text{or} \\
\label{eq:fullyEncryptedDynamicZ}
\ct(z(k+1)) &= \ct(z(k)) \oplus \ct(\lfloor s \Delta t \rceil) \otimes \ct( \lfloor s y(k)  \rceil). 
\end{align}

\vspace{-3mm}
\begin{table}[h]
    \centering
        \caption{Summary of encrypted benchmark controllers.}
    \label{tab:benchmarkControllers}
    \begin{tabular}{lcc}
    \toprule
      Controller type   &  Partially encrypted & Fully encrypted  \\
      \midrule 
       State feedback  & via \eqref{eq:partiallyEncryptedKy} & via \eqref{eq:fullyEncryptedKy} \\
       PI control & via \eqref{eq:partiallyEncryptedDynamicU} and \eqref{eq:partiallyEncryptedDynamicZ}  & via \eqref{eq:fullyEncryptedDynamicU} and \eqref{eq:fullyEncryptedDynamicZ} \\
       \bottomrule
    \end{tabular}

\end{table}

\vspace{-3mm}
\subsection{Benchmark system}
\label{subsec:system}

In summary, we will consider the encrypted controllers in Table~\ref{tab:benchmarkControllers} for our tutorial and benchmark.
These controllers will be applied to the benchmark system from \cite[Exmp.~9]{aastrom2000benchmark} with $\omega_0=10$. 
A time-discretization with the sampling time $\Delta t=1$ results in the state space model 
\begin{align}
\nonumber
    \xb(k+1) \! &=\! \begin{pmatrix}
        -0.27 & \hspace{-0.4em} \blind{-}0.24 & \hspace{-0.4em}\blind{-}0.08 \\
        -0.20 & \hspace{-0.4em} -0.35 & \hspace{-0.4em} -0.17 \\
        \blind{-}0.22  & \hspace{-0.4em} -0.02 & \hspace{-0.4em} \blind{-}0.36
    \end{pmatrix} \xb(k) \!+\! \begin{pmatrix}
       -0.05  \\
        \blind{-}0.11 \\
        \blind{-}0.41
    \end{pmatrix} \!u(k) \\
    \label{eq:sysDynamics}
    y(k)&=\begin{pmatrix}
        0 & 0 & 1.56
    \end{pmatrix} \xb(k).
\end{align}
For these dynamics, stabilizing state feedback of the form \eqref{eq:uKx} is given for, e.g., 
\begin{equation}
\label{eq:KxNumbers}
 \Kb= \begin{pmatrix}
      -0.07 & 0.06 & -0.12
 \end{pmatrix}.
\end{equation}
Stabilizing PI control results, e.g., for the controller gains
\begin{equation}
\label{eq:KpiNumbers}
K_{\text{p}}=-0.5 \quad \text{and} \quad  K_{\text{i}}=-0.75.
\end{equation}
The system dynamics~\eqref{eq:sysDynamics}and the controller parameters~\eqref{eq:KxNumbers} and \eqref{eq:KpiNumbers} will be used for the following encrypted implementation and the numerical studies in Section~\ref{sec:numericalStudies}.
For all experiments, we consider the initial state $\xb_0=(\,
    10 \,\,\, 10 \,\,\, 10\,)^\top$.

\section{Encrypted realizations via toolbox}
\label{sec:implementation}

Next, we illustrate how to implement the encrypted controllers from Section~\ref{subsec:benchmarkcontrollers} for the benchmark system in Section~\ref{subsec:system} using the cryptosystems from Section~\ref{sec:basics}. 
To this end, we implemented a Matlab Toolbox which is available on our GitHub repository\footnote{see \href{https://github.com/Control-and-Cyberphysical-Systems/ECC-Tutorial}{https://github.com/Control-and-Cyberphysical-Systems/ECC-Tutorial}\!}. 
The code supports this tutorial by providing additional details regarding the cryptosystems and allowing for basic experiments. 
We recommend to study the exemplary controller implementations provided in the toolbox along with the following text-based presentations.

Before presenting the encrypted controllers, we make three comments to clarify our approach. First, our focus lies on controller evaluations rather than actually networked implementations, which we consequently omit for the sake of simplicity. 
Second, our toolbox is designed for educational purposes and mainly offers accessibility as well as simplicity (but not performance) compared to state-of-the-art implementations such as \cite{al2022openfhe}. Third, some of the following parameter choices serve illustrative purposes but may not satisfy real-world security demands.

\subsection{Realizations using Paillier}
\label{subsec:RealPaillier}

To setup the Paillier cryptosystem, we specify the security parameter $\lambda$ (here bit-length of $p_i$) and call $\mathsf{KeyGen}()$:
\begin{lstlisting}[style=matstyle] 
digits(500)   % increase precision
lambda = 128; % security parameter
% public and secret key with bitlength `\!\color{matgreen}{$\lambda$}`
[pk,sk] = KeyGen(lambda) 
\end{lstlisting}
Note that \mattext{digits(500)} increases the precision used by the variable precision arithmetic \mattext{vpa()} to $500$ digits. For cryptographic purposes, precise integer arithmetic is critical for correctness. Our implementation uses $\lambda=128$ ($1024$ is actually recommended for proper security), which results in ciphertext magnitudes of $2^{256}$. Hence, an implementation using standard \mattext{double} data types would simply fail.
Next, given a message $x\in\R$, one can encrypt and decrypt using 
\begin{lstlisting}[style=matstyle] 
x = vpa(1.2345);   % define vpa variable
s = vpa(10)^3;     % scaling factor
z = Ecd(x,s,pk);   % map to `\!\color{matgreen}{$\Z_{\pk}$ via \eqref{eq:encoding}}`
ct_z = Enc(z,pk);  % encrypt via `\!\color{matgreen}{\eqref{eq:paillierencdec}}`
\end{lstlisting}
while the homomorphisms given $\ct(z)$ and $z_2\in\Z_{\pk}$ are
\begin{lstlisting}[style=matstyle] 
ct_z3 = ct_z+ct_z; % encrypted addition
ct_z4 = ct_z*z2;   % public multiplication
\end{lstlisting}
For convenience, ``\texttt{+}'' and ``\texttt{*}'' are overloaded to intuitively implement \eqref{eq:paillierhom} and~\eqref{eq:paillierMul}, respectively.
Furthermore, different input types are handled automatically.
Correctness of the decryption is ensured
as long as $z_3,z_4\in\Z_{\pk}$. For instance, we obtain
\begin{lstlisting}[style=matstyle] 
z3 = Dec(ct_z3,sk); % decryption via `\!\color{matgreen}{\eqref{eq:paillierencdec}}`
Dcd(z3,s,pk)        % decoding via `\!\color{matgreen}{\eqref{eq:decoding}}`
    ans = 2.468
\end{lstlisting}
as expected.
Now, due to the overloaded operators, an implementation of partially encrypted state feedback and PI control is compactly realized as follows. 

\textit{State feedback.} We initialize the system matrices \mattext{A,B,} and \mattext{C} according to~\eqref{eq:sysDynamics} and perform a key generation as above. 
The initial state is set to \mattext{x =[10; 10; 10]} and the feedback gain is specified as in \eqref{eq:KxNumbers}. Both quantities are first encoded and $x$ is subsequently encrypted via
\begin{lstlisting}[style=matstyle] 
 s = vpa(10)^3;
zx = Ecd(vpa(x),s,pk);
zK = Ecd(vpa(K),s,pk);
ct_x = Enc(zx,pk);
\end{lstlisting}
Here, \mattext{Ecd($\cdot$)} and \mattext{Enc($\cdot$)} perform element-wise 
encodings and encryptions, respectively.
Finally, the control loop is simulated, where the main routine is as follows:
\begin{lstlisting}[style=smallerstyle]
ct_u = zK*ct_x;                     % @ cloud
  zu = Dec(ct_u,sk);                % @ actuator
u(k) = Dcd(zu,s^2,pk);              % @ actuator
 x(:,k+1) = A*x(:,k)+B*u(k);        %   system
zx(:,k+1) = Ecd(vpa(x(:,k+1)),s,pk);% @ sensor
ct_x = Enc(zx(:,k+1),pk);           % @ sensor
\end{lstlisting}
Here, indicated by the input dimensions and data types, \mattext{zK*ct\_x} executes \eqref{eq:partiallyEncryptedKy} automatically. 
The result is then decrypted at the actuator and decoded according to~\eqref{eq:uKyInteger}.
Subsequently, the actuator applies the control input \mattext{u(k)}, the system state evolves, and the state is encoded and encrypted at the sensor.

\textit{PI control.} Next, we consider the PI controller~\eqref{eq:scalarPI} with the parameters~\eqref{eq:KpiNumbers}. 
Since many implementation steps are similar to those for state feedback, we concentrate on the steps that differ.
We initially set the controller state \mattext{xc} to zero and encode the controller parameters, $\Delta t$, and \mattext{xc}. Then, the initial state is encrypted via:
\begin{lstlisting}[style=matstyle] 
 s  = vpa(10)^3;
zdt = Ecd(vpa(dt),s,pk);
zKi = Ecd(vpa(Ki),s,pk);
zKp = Ecd(vpa(Kp),s^2,pk);
zxc = Ecd(vpa(xc),s^2,pk);
ct_xc = Enc(zxc,pk);
\end{lstlisting}
Finally, the main routine becomes
\begin{lstlisting}[style=smallerstyle]
y(k) = C*x(:,k);            %   system output
zy = Ecd(vpa(y(k)),s,pk);   % @ sensor
ct_y = Enc(zy,pk);          % @ sensor
ct_u = zKi*ct_xc+zKp*ct_y;  % @ cloud
ct_xc = ct_xc+zdt*ct_y;     % @ cloud
zu = Dec(ct_u,sk);          % @ actuator
u(k) = Dcd(zu,s^3,pk);      % @ actuator
x(:,k+1) = A*x(:,k)+B*u(k); %   system state
\end{lstlisting}
according to \eqref{eq:partiallyEncryptedDynamicU}, \eqref{eq:partiallyEncryptedDynamicZ}, and~\eqref{eq:sysDynamics}.

\subsection{Realizations using GSW}
Now, for implementations based on the GSW scheme, we begin with $\mathsf{Setup}(\lambda,q)$ and $\mathsf{KeyGen()}$, i.e., 
\begin{lstlisting}[style=matstyle] 
digits(100)          % increase precision
B = vpa(10);         % noise bound
N = 4;               % key dimension  
q = vpa(10)^20;      % ciphertext space `\!\color{matgreen}{$\Z_q$}`
setup = Setup(N,q,B); 
sk = KeyGen(setup);
\end{lstlisting}
Here, \mattext{Setup(N,q,B)} mainly collects parameters while \mattext{KeyGen(setup)} generates the secret key $\sk$ (via sampling). 
At this point, we note that selecting parameters of an LWE instance such that a certain security $\lambda$ is ensured, is a non-trivial problem. Still, for a given noise distribution and $q$, one can use the LWE estimator \cite{albrecht2015concrete} to find a suitable $N$ such that a security of $\lambda$ is ensured. Here, we simply use $N=4$ (which would be far too small for a secure implementation).
Remarkably, in contrast
to Paillier, where the ciphertext and message spaces are directly related to $\lambda$, LWE-based schemes allow selecting $q$, suitable for a given computation, independently of $\lambda$ due to $N$.

Similarly to before, one can encrypt, perform encrypted computations, and decrypt. However, there exist two types of ciphertexts (LWE and GSW) and encrypted multiplications are possible as follows:
\begin{lstlisting}[style = matstyle]
s = vpa(10)^3;
zx = Ecd(vpa(1.2345),s,q);
zy = Ecd(vpa(4.5678),s,q);
ct_x = LWE(setup,zx,sk);     % LWE encryption
ct_y = GSW(setup,zy,sk);     % GSW encryption
ct_xy = ct_x*ct_y;
xy = Dcd(Dec(ct_xy,sk),s^2,q)
    ans = 5.646
\end{lstlisting}
Again, the operators are overloaded, e.g., ``\texttt{*}'' performs \eqref{eq:LWEtoGSW}. One important observation 
 is that it is non-trivial to transform an LWE into a GSW ciphertext. Thus, if the multiplicative depth is larger than one, \eqref{eq:GSWtoGSW} should be used. However, this is neither required in static state feedback nor in PI control.

\textit{State feedback.} Based on the previous explanations regarding a Paillier-based evaluation, we have already seen many of the syntactically (almost) equal steps in the GSW-based implementation. Thus, we will again only point out differences. Namely, one needs to additionally encrypt \mattext{zK} using a GSW ciphertext \mattext{ct\_K=GSW(setup,zK,sk)}. Here, \mattext{GSW()} automatically encrypts \mattext{zK} element-wise.
Second, the main routine differs by the fully encrypted controller evaluation \mattext{ct\_u=ct\_K*ct\_x}, which implements \eqref{eq:fullyEncryptedKy} as expected.

\textit{PI control.} In comparison to the Paillier case, we additionally compute encryptions of $\round{s\Delta t}$,$\round{s\Delta K_{\text{i}}}$, and $\round{s^2K_{\text{p}}}$ 
\begin{lstlisting}[style = matstyle]
ct_dt = GSW(setup,zdt,sk);
ct_Ki = GSW(setup,zKi,sk);
ct_Kp = GSW(setup,zKp,sk);
\end{lstlisting}
which allows us to evaluate the PI controller in a fully encrypted manner. In this context, the controller evaluation via \eqref{eq:fullyEncryptedDynamicU} and \eqref{eq:fullyEncryptedDynamicZ} in the main routine changes to
\begin{lstlisting}[style = matstyle]
ct_u  = ct_Ki*ct_xc+ct_Kp*ct_y;
ct_xc = ct_xc+ct_dt*ct_e;  
\end{lstlisting}
where \mattext{ct\_xc,ct\_y,ct\_u,ct\_e} are LWE ciphertexts.

\subsection{Realizations using CKKS}
Lastly, we focus on CKKS. Because CKKS is also based on (R)LWE, the $\mathsf{Setup}()$ and $\mathsf{KeyGen}()$ algorithms are quite similar, and the consideration regarding the security parameter $\lambda$ apply here as well. In particular, one can use the following code:
\begin{lstlisting}[style = matstyle]
digits(100)
s = vpa(10)^3;      % scaling factor
q = vpa(10)^15;     % base modulus
L = 2;              % number of levels
N = 4;              % key dimension
setup = Setup(N,q,s,L);
[pk,sk,evk] = KeyGen(setup);
\end{lstlisting}
However, due to the operation of CKKS, one needs to specify the level \mattext{L} directly (this happened indirectly via $q$ in GSW).
Moreover, the noise is automatically selected and the \mattext{KeyGen(setup)} generates $\pk$, $\sk$, and the evaluation key $\evk$, which is required for encrypted multiplications.

Now, we need to encode a message into the cryptosystem's message space before we can encrypt it. In this context, RLWE schemes are special in contrast to most cryptosystems because they operate over $\Rc_q$ instead of $\Zc_q$ (or $\Z_q$).
Here, the encoding becomes
\begin{lstlisting}[style = matstyle]
x = vpa(1.2345);
z = Ecd(setup,x,s,'type','scalar');
\end{lstlisting}
which results in the polynomial
$$
    z(X) = 0\!\cdot\! X^{3} + 0\!\cdot\!X^{2} + 0\!\cdot\!X^{1} + 1234\!\cdot\! X^{0} \in\Rc_q.
$$
For simplicity, we use the option \mattext{scalar} such that the encoding performs an element-wise scaling and rounding as in \eqref{eq:encoding} and automatically pads the message with zeros from the left. 
Alternatively, there exists a \mattext{packed} encoding in which a DFT-like transformation is computed that enables coefficient-wise (simultaneous) additions and multiplications. 

Then, the encryption is \mattext{ct\_x = Enc(setup,z,pk)}. In contrast
to Paillier and GSW, this ciphertext allows for $L$ subsequent multiplications. Suppose \mattext{ct\_x,ct\_y} are given with a level of $L=2$, one can, e.g., compute the (multivariate) polynomial
\begin{lstlisting}[style = matstyle]
ct_poly = ct_x^4+ct_y^2*(ct_x-ct_y)+ct_y;
\end{lstlisting}
which has a multiplicative depth of $2$. In addition to ``\mattext{+}'' and ``\mattext{*}'', also ``\mattext{\textasciicircum}'' is overloaded with the corresponding homomorphic operations. After each multiplication, the ciphertext level is automatically reduced by one.
The decryption and decoding then amount to \mattext{p = Dcd(Dec(ct\_poly,sk))} with the polynomial $p\in\Rc_q$ which contains the desired result in the coefficient related to $X^0$.

\textit{State feedback \& PI control.} First, we proceed analogously to the GSW implementation but replace the encodings and encryptions by
\mattext{Ecd(setup,$\cdot$,s,'type','scalar')} and \mattext{Enc(setup,$\cdot$,pk)}, respectively, where both act element-wise. 
Furthermore, we consider the slightly different decodings (as explained above).
Apart from that,
the implementation is identical to the GSW implementation. We finally note that, for state feedback, $L=1$ suffices, whereas PI control requires $L=2$.

\section{Numerical experiments}
\label{sec:numericalStudies}

\begin{figure}[h]
    \centering
    \includegraphics[width=\linewidth,trim=4cm 11.5cm 4cm 11cm, clip]{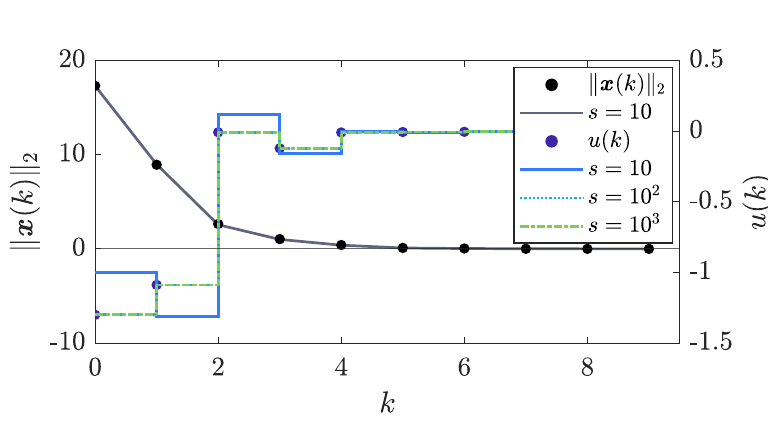}
    \caption{State (left) and input (right) trajectories for encrypted state feedback and varying accuracy (curves with dots refer to plaintext values; colors refer to different scaling factors).}
    \label{fig:stateresult}
  \end{figure}

\begin{figure}[h]
    \centering
\includegraphics[width=\linewidth,trim=4cm 11.5cm 4cm 11cm, clip]{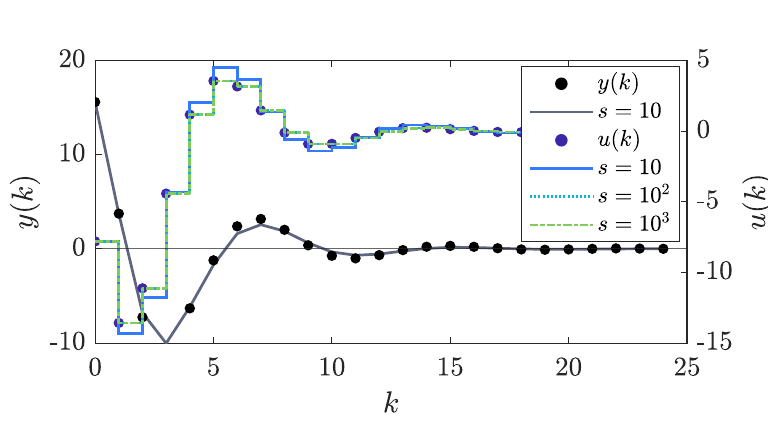}
    \caption{Outputs (left) and inputs (right) for encrypted PI control and varying accuracy (with line styles as in~Fig.~\ref{fig:stateresult}).}
    \label{fig:PIresult}
\end{figure}

In this section, we briefly analyze the closed-loop performance of the encrypted control schemes for varying accuracy (depending on the scaling factor $s$).
Since the closed-loop behavior for our numerical experiments turns out to be be almost identical independent of the cryptosystem in use, we only discuss the numerical results obtained for Paillier-based controller encryptions.
The corresponding realizations discussed in Section~\ref{subsec:RealPaillier} lead to the input, state, and output trajectories in Figures \ref{fig:stateresult} and \ref{fig:PIresult}. As apparent, the states and outputs converge to the desired (zero) setpoints. Furthermore, we observe that the effect of varying scaling factors (i.e., $s \in \{10,10^2,10^3\}$) is minor for the studied control system. In fact, while the input trajectories slightly vary, the state and output trajectories are almost identical to their plaintext counterparts for every considered $s$.


The numerical experiments allow for another noteworthy observation. In fact, while the multiplicative depth is finite for both schemes, the PI controller builds on the recursion~\eqref{eq:recursionPI}, which results in an unbounded number of encrypted additions (for an unlimited runtime). Hence, one might expect an overflow resulting from encrpyted additions for the GSW- and CKKS-based implementations after some time. However, due to the stabilized closed-loop,  errors are attenuated.
This observation is formalized in \cite{kim2020comprehensive}.

\section{Summary and Outlook}
\label{sec:outlook}

We provide a code-driven tutorial on encrypted control based on a custom Matlab toolbox developed by the authors.
The tutorial illustrates the encrypted implementation of state feedback and PI control. Essential concepts and limitations are discussed along with numerical experiments.

The introductory tutorial can be extended in various directions. For instance, we may consider more functionalities, performance optimizations, latency, or the integration of an actual cloud.

\end{document}